\documentclass[preprint,12pt]{elsarticle}

\usepackage{amssymb}
\usepackage{amsmath}
\usepackage{graphicx}
\usepackage{dcolumn}	
\usepackage{bm}			
\usepackage{color}
\usepackage{natbib}

\journal{Physica A}

\begin{document}

\begin{frontmatter}
\title{Networks with time structure from time series}
\author[a1]{Tomomichi Nakamura}
\ead{tomo@sim.u-hyogo.ac.jp}
\author[a2]{Toshihiro Tanizawa}
\ead{tanizawa@ee.kochi-ct.ac.jp}
\address[a1]{Graduate School of Simulation Studies, University of Hyogo,\\
7-1-28 Minatojima-minamimachi, Chuo-ku, Kobe, Hyogo 650-0047, Japan}
\address[a2]{Kochi National College of Technology, 200-1 Monobe-Otsu, Nankoku,\\
Kochi 783-8508, Japan}
\begin{abstract}
We propose a method of constructing a network,
in which its time structure is directly incorporated,
based on a deterministic model from a time series.
To construct such a network, we transform a linear model containing
terms with different time delays into network topology.
The terms in the model are translated into temporal nodes of the network.
On each link connecting these nodes, we assign a positive real number 
representing the strength of relationship, or the ``distance,''
between nodes specified by the parameters of the model.
The method is demonstrated by a known system and applied to 
two actual time series.
\end{abstract}
\begin{keyword}
time series modelling; 
networks;
minimum spanning trees
\PACS{05.45.Tp, 89.20.Ff, 89.75.Fb, 89.75.Hc}
\end{keyword}
\end{frontmatter}

\section{Introduction}
\label{sec:introduction}

Understanding the complex features of various dynamical systems 
in the real world continues to be a crucial challenge across 
the physical and natural sciences.
To better understand such complicated interactions it is useful  
to first transform the system into a new frame of reference.
This is the approach we follow here. 
In this paper, we describe a method to transform a linear model
built from a time series representing a
(possibly nonlinear, possibly stochastic)
dynamical system into a network which is capable of representing 
the time structure of the model to explore undiscovered structure 
of the original dynamical system.
This method enables us to investigate the dynamics of the time series
from a network perspective and allows us to explore the globally interrelated 
structure and the hierarchy.
The network transformation provides us with a new representation 
of a generic model of the original dynamics.
By examining the topology of the network, 
we can directly obtain better understanding of the structure and
the characteristics of the phenomenon represented by the time series modeling.
Connectivity between different features of the deterministic model structure 
can now be understood in greater detail than before thanks to 
this new toolbox from complex network theory.

Over the past decade it has become clear that networks as typified by
``complex networks'' have a vast range of applicability and give a new perspective 
on various problems in the real world~\cite{Watts-etal:network98, barabasi99, albert02, newman03}. 
In particular, there have been several recent works to combine time series 
with networks,
such as networks through the correlation strength~\cite{Yamasaki-etal:EL_Nino},
recurrence networks~\cite{Marwan-etal:Recurrence_network},
cycle networks for pseudo-periodic time series~\cite{Zhang-Small:complex_network,Xu-eatl:network08}
and the horizontal visibility algorithm~\cite{Luque-eatl:network09}.
These approaches are proven to be effective in understanding
complicated and entangled structure of systems~\cite{%
Yamasaki-etal:EL_Nino, Small-etal:complex_network09, Luque-eatl:network11}.
However, these works still do not address one essential ingredient.
Although systems in the real world have various different time delay or 
feedback effects in principle, such effects,
which we refer to as ``time structure,''
cannot be directly treated or set in the network by current approaches.
Furthermore, the network structures from time series
in the existing approaches may not always be sufficient to capture 
the salient features of the underlying global interrelation structure,
because these structures are essentially built on local information
such as the temporal structures on an embedding space and 
correlations (similarities) between each pair of time series among 
many~\cite{Yamasaki-etal:EL_Nino, Marwan-etal:Recurrence_network, Small-etal:complex_network09}. 
In addition, ``no similarity'' is not equivalent to ``no correlation.'' 
Even when two signals are not similar, these systems can still 
have some kind of correlated structures.
Under these circumstances, it is crucial to find a method to translate
directly the features of dynamics into network topology.

In this paper, to unravel entangled deterministic dynamical structures,
we propose a method of constructing a network based on deterministic model structure 
from a given time series.
Time structure is directly set in the network and the network fully reflects 
the global structure of interrelation and the hierarchy of the components of the model. 
This is the first attempt to represent a generic deterministic model
as a network in order to understand the deterministic dynamics 
of observed data with its time structure fully incorporated.

It is usually difficult to extract the hidden or underlying nature
of the structure of a given complicated time dependent phenomenon
by directly examining it.
However, it is relatively easy to obtain a time series data and 
build a model from the data.
In principle, we can expect that the time series model reflects underlying features 
of the time series such as entangled time structures in varying degrees 
and therefore the model can be a good starting point for unravelling the time structure 
of the system.
It should be emphasized that the method for transforming the time series model 
into network topology presented in this paper offers an approach completely 
different from existing ones to clearly visualise the hidden time structure
of the system.

This paper is organized as follows.
In Section~\ref{sec:Algorithm}, the network construction
algorithm from a given time series proposed in this paper is described.
To demonstrate a concrete example of this algorithm,
it is applied to a simple ``toy model'' in Section~\ref{sec:numerical example}.
The effectiveness of the algorithm is shown in Section~\ref{sec:application},
where the method is applied to two actual time series: 
annual sunspot numbers and a microtremor data set from an earthquake.
We summarise the results in Section~\ref{sec:summary}.

\section{Algorithm for transforming a linear model to a network}
\label{sec:Algorithm}

To construct networks with time structure,
we have to find the underlying time delays in time series
as necessary elements and translate them into a network 
for visualization.
An effective first step is the time series
modelling~\cite{Judd-Mees:non-uniform_embedding,%
Small-Judd:RAR,Tong-Lim:piecewise_linear_model}.
To obtain the precise interrelation of terms with time delays or
the essential linear structures included in time series,
we use an information theoretic reduction of linear models,
the reduced autoregressive~(RAR)
model~\cite{Judd-Mees:non-uniform_embedding,Small-Judd:RAR}.
Unlike a standard AR model, RAR models include only terms that contribute
significantly to the model, as assessed by an information criterion.
A standard AR model is built up from terms with unit time delay,
though some terms (some time delays) might not be necessary
to capture the essential feature of phenomena.
In contrast, RAR models include only terms that contribute significantly 
to the model as assessed by an information criterion\footnote{%
Although we consider that the RAR model is the best approach for the proposed method, 
we can construct a network using a standard AR model.}.

The RAR model has proven to be effective
in modelling both linear and nonlinear
dynamics~\cite{Judd-Mees:non-uniform_embedding,Small-Judd:RAR},
and this model provides a generic method to locally linearize
the system under consideration in a way which would be applicable
to any smooth vector field,
even though the underlying system is not linear.
More details about RAR models can be found in
the literature~\cite{Judd-Mees:non-uniform_embedding,Small-Judd:RAR}.

\subsection{Building an RAR model}
\label{sec:building RAR model}

The method described in this paper is composed of two steps:
(i)~building an RAR model from a given time series
and (ii)~constructing a network from the model.
We first give a brief review of the RAR model.

Given a time series $ \{x_t\}_{t=1}^n $ of $ n $ observations,
an RAR model with the largest time delay~$ w $ can be expressed by 
\begin{equation}
    x(t) = a_{0} + a_{1} x(t - l_{1}) + a_{2} x(t - l_{2}) + \dots + a_{w} x(t - l_{w}) + \varepsilon(t),
    \label{eq:RAR}
\end{equation}
where $ a_i \; (i = 0, 1, 2 \dots, w) $ are unknown parameters, 
and $ \varepsilon(t) $ is assumed to be independent and identically
distributed Gaussian random variables, which are interpreted as a fitting error.
The parameters~$ a_i $ are chosen to minimize the sum of squares of 
the fitting errors.

Among various information criteria used to find the best (optimal)
model~\cite{Nakamura-etal:information_criteria},
we employ the Schwartz information criterion~(SIC)~\cite{Schwarz:SIC}.
The SIC formula is defined by
    \begin{eqnarray}
        \mathrm{SIC}(k) = n \ln \frac{ {\bf e}^T {\bf e} }{n} + k \ln n,
    \label{eq:SIC_eq}
    \end{eqnarray}
where $ n $ is the number of data points, $ k $ is the model size
and $ \bf e $ is the fitting errors\footnote{The SIC is also known 
    as the Bayesian Information Criterion~(BIC) and description length 
    proposed by Rissanen has essentially the same formula~\cite{Nakamura-etal:information_criteria}.
    There is a variety of information criteria~(IC), 
    each with a different background~\cite{Nakamura-etal:information_criteria}.
    Although different IC may build different model, 
    the proposed method can equally construct the corresponding network.}.

In building RAR models, we need to select necessary terms.
Firstly, many candidate basis functions are
prepared in the form of a dictionary.
Next we select as many basis functions that can extract the peculiarity of the time series
as possible according to various model selection methods~\cite{Judd-Mees:non-uniform_embedding}.

All of these model selection methods have a common difficulty, which is
to identify the globally optimal model, due to trapping into local
minima~\cite{Nakamura-etal:information_criteria}.
    For finding the truly optimal model,
    all possible combinations of basis functions have to be calculated (an exhaustive search).
    Although an exhaustive search for a dictionary smaller than approximately 30 is usually manageable,
    the calculation for a large dictionary costs enormous computational time and is unrealistic,
    as performing an exhaustive search is expected to be an NP-hard problem~\cite{Judd-Mees:non-uniform_embedding}.
    It should be noted, however, that
    we can always resort to other methods for approximately finding 
    the global minimum~\cite{Judd-Mees:non-uniform_embedding},
    when exhaustive search is computationally impossible.
    The models obtained by selection algorithms are nearly optimal.
    
To avoid this complexity caused by near optimality,
we use exhaustive search in this paper in finding the optimal model that gives the minimum value 
of SIC~\cite{Nakamura-etal:information_criteria}.
For verification,
we also investigated the network structure of nearly optimal models
and found that the core structure of the network built up
    from the optimal model is preserved
    in these nearly optimal models
    apart from the other ``leaf'' nodes around the core structure.

\subsection{Constructing a network}
\label{sec:construction of networks}

After obtaining a model representing the time series under consideration,
we transform the model into a directed network
(i)~by representing each term~$ x(t) $ at time~$ t $ in the model by a node labelled by the time
and (ii)~by drawing an arrow directed from a node~$ x(t-i) $ 
to the node~$ x(t) $, where the time delay term $ x(t-i) $ appears 
in the RAR model for the expression of $ x(t) $.  
This arrow represents the influence of $ x(t-i) $ on $ x(t) $ with 
time delay~$ i $.
It seems reasonable to assume that
a large absolute value of the parameter $ a_i $ represents 
the large influence of $ x(t-i) $ on $ x(t) $.  
We treat the influence as a ``distance'' between nodes~$ x(t) $ 
and $ x(t-i) $ using $ a_i $ on the network space;
the larger the absolute value of $ a_i $,
the shorter the distance between $ x(t) $ and $ x(t-i) $.
Since there have been no previous attempts to translate
the values of the parameters into the network space,
we introduce the following simple distance based on
elementary linear algebra.

We should mention the constant parameter~$ a_0 $ and dynamical noise~$ \varepsilon(t) $
in the RAR model.
They surely play important roles in actual dynamics
and both of them should be included in constructing a model.
However, these terms contain no time information.
Since our purpose is to unravel entangled time structure of deterministic dynamics, 
we do not take them into account in defining the distance
between terms with time delay in constructing a network.

Equation (\ref{eq:RAR}) can be interpreted as a linear combination of
the set of linearly independent ``unit vectors,''
\( x(t - l_{1}), x(t - l_{2}), \dots, x(t - l_{w}) \),
with the coefficients, $a_1, a_2, \dots, a_w$.
By this interpretation, we introduce the ``angle''~$ \theta_i $ 
between the directions of $ x(t) $ and $ x(t - i) $ as
\begin{equation}
    \theta_{i} \equiv \arccos \left( \frac{a_i}{\sqrt{a_1^2 + a_2^2 
           + \dots + a_w^2}} \right).
    \label{eq:angle}
\end{equation}
The distance we introduce should have following properties.
Firstly, when vectors~$ x(t) $ and $ x(t-i) $ are in the same direction,
the angle $ \theta_i $ becomes $ 0 \text{ or } \pi $
and the distance~$ d_i $ should be $ 0 $.
We expect the analyticity of the ``distance'' around $ \theta = 0 $ 
and put $d_i \approx \theta_i$ in this case.
Secondly, when the vectors $ x(t) $ and $ x(t-i) $ are perpendicular ($ a_i = 0 $), 
the angle~$ \theta_i $ becomes $ \pi/2 $ and the distance~$ d_i $ 
should be infinity. 
Thus $ d_i $ should be inversely proportional to $ \cos \theta_i $.
Finally, the distance must always be a positive real number.
Regarding all these requirements,
we define the distance~$ d_i $ between the nodes $ x(t) $ and $ x(t-i) $ as
\begin{equation}
    d_i \equiv \left| \tan \theta_{i} \right|.
    \label{eq:distance}
\end{equation}

We consider that the concept of ``distance'' introduced here depends on 
the nature of the system, and other distances such as the inverse of 
the parameters may be justified in some situations.
However, we consider that this distance is appropriate for most cases of linear models,
because it reflects the overall balance of the size of parameters in the model.
Note that the proposed method is independent of the definition of the distance.

According to Eq.~(\ref{eq:RAR}), the nodes contained in a model
are directly connected to $ x(t) $.
The distance calculated by Eq.~(\ref{eq:distance}) is referred to
as the direct distance~({\it DD}).
A pair of nodes, however, can be connected indirectly via some other nodes.
The sum of all distances through the path between these two nodes
is referred to as the indirect distance ({\it ID}).
In a network we sometimes find a path with a shorter {\it ID} than the {\it DD}.
In such a case, we can consider that the indirect path that gives 
the shortest length is the path on which the information is passed through
most effectively and that these two nodes are essentially connected
through the shortest {\it ID} path.
We therefore treat the collection of the paths that have the shortest length
for any given node pairs as the network of the system.
The network constructed in this way reveals the underlying hierarchical 
structure of the linear model and enables us to know whether the influence 
of a term may come through other terms.

\section{Numerical example}
\label{sec:numerical example}

We demonstrate the application of our algorithm and
confirm our theoretical argument with a simple example.
We begin by the following (artificial) RAR model:
\begin{equation}
    x(t) = 1.01 \; x(t-1) - 0.61 \; x(t-3) + 0.11 \; x(t-6) + \varepsilon(t).
    \label{eq:simple example}
\end{equation}

\begin{figure}[btp]
\begin{center}
\includegraphics[clip, width=8.0cm]{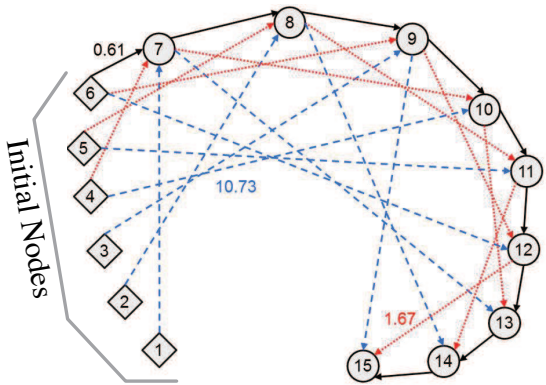}
\end{center}
\caption{(Colour online)~The whole linkage
 information contained in the network representation of the first 15 terms
 (that is, $ x(1) $ to $ x(15) $) represented by the RAR model,
 Eq.~(\ref{eq:simple example}), where the numbers on the nodes
 are $ t $ of $ x(t) $ in Eq.~(\ref{eq:simple example}).
 The first six nodes represented by squares are the initial nodes.
 The numbers on the arrows are the direct distances between nodes.
 The positions of the nodes are irrelevant (only the
 topology is important) and
 the length of the arrows also does not represent the actual scale.
 From this figure, we can clearly see that there are cases
 where an indirect distance between a pair of nodes is
 much shorter than the direct distance.
 For example, the indirect distance between node $ x(9) $ and
 $ x(15) $ via node $ x(12) $ is $ 1.67 + 1.67 = 3.34 $, 
 while their direct distance is $ 10.73 $.}
 \label{fig:network for simple example}
\end{figure}

From the definitions of Eqs.~
(\ref{eq:angle}) and (\ref{eq:distance}), we obtain
the direct distances ({\it DD}\/s) between the nodes as
$ \vec{d} = \left( 0.6137, \; 1.6655, \; 10.7265 \right) $.
In Fig.~\ref{fig:network for simple example}, we show the overall linkage
of the network constructed from Eq.~(\ref{eq:simple example}) within
the time interval from 1 to 15.
%

The distance between nodes represents the magnitude of influence 
from the other nodes. According to the model, 
the nodes~$ x(t-1) $, $ x(t-3) $ and $ x(t-6) $
are directly connected with the node $ x(t) $.
For example, the nodes $ x(6) $, $ x(4) $ and $ x(1) $ are directly connected 
with the node $ x(7) $ in Fig.~\ref{fig:network for simple example}.
As mentioned in the previous Section, however, the {\it DD} is not always the shortest.
For example, the optimal~(shortest) path from $ x(9) $ to $ x(15) $
is the one via $ x(12) $, while the last node~$ x(15) $ is directly 
connected with $ x(9) $ as Fig.~\ref{fig:network for simple example} shows.
Since both of the {\it DD}\/s from $ x(9) $ to $ x(12) $
and from $ x(12) $ to $ x(15) $ are 1.67,
the {\it ID} from $ x(9) $ to $ x(15) $ is the summation, 
$ 1.67 + 1.67 = 3.34 $, which is much shorter than
the {\it DD} from $ x(9) $ to $ x(15) $, $ 10.73 $,
We can thus conclude that the most significant influence of 
the term $ x(t-6) $ to $ x(t) $ is not the direct one but that 
comes through the term $ x(t-3) $.

%
\begin{figure}[btp]
\begin{center}
\includegraphics[clip, width=7.0cm]{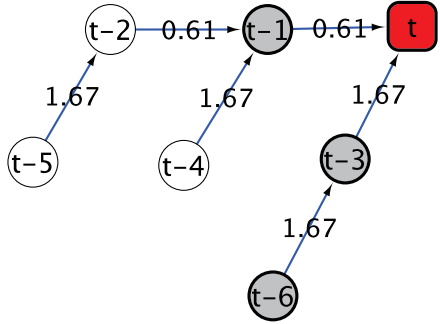}
\end{center}
\caption{(Colour online)~The optimal path network constructed from the model,
         Eq.~(\ref{eq:simple example}). 
         The nodes shown are from $ x(t) $ to the one with the largest
         time delay, $ x(t-6) $.
         Note that the node $ x(t-6) $ is not directly connected to $ x(t) $ 
         but connected through $ x(t-3) $.
         The gray colour corresponds to terms included in the model
         and the white colour to terms within the largest time delay
         but not included in the model, respectively.
         The numbers on the arrows are the direct distances between the nodes.
         We show all terms within the largest time delay 
         to show the flow of time in the model without gaps
         and the structure more clearly.}
    \label{fig:mst network for simple example}
\end{figure}

To show this situation more clearly,
we show in Fig.~\ref{fig:mst network for simple example}
the set of most optimal paths to node $ x(t) $ from the nodes within the model 
(between $ x(t) $ and $ x(t-6) $).
Figure~\ref{fig:mst network for simple example} shows that
the node $ x(t) $ is directly connected with $ x(t-1) $ and $ x(t-3) $,
and the connection from $ x(t-6) $ is indirect via $ x(t-3) $.
This is an outcome of the interplay between the sizes of the parameters
caused by the interrelation and hierarchy between terms of the model, Eq.~(\ref{eq:simple example}).
In this way, the network topology constructed by the present method reveals the time structure.
Note that we cannot extract this information by simply examining
Eq.~(\ref{eq:simple example}).
This is the manifestation of the underlying structure of dynamical systems 
through the network representation.

\section{Application to real world data}
\label{sec:application}

Based on the result of these computational studies,
we apply the proposed method to two real world data:
(i)~annual sunspot numbers and (ii)~microtremor data.

\begin{figure}[btp]
\begin{center}
\includegraphics[clip, width=10.0cm]{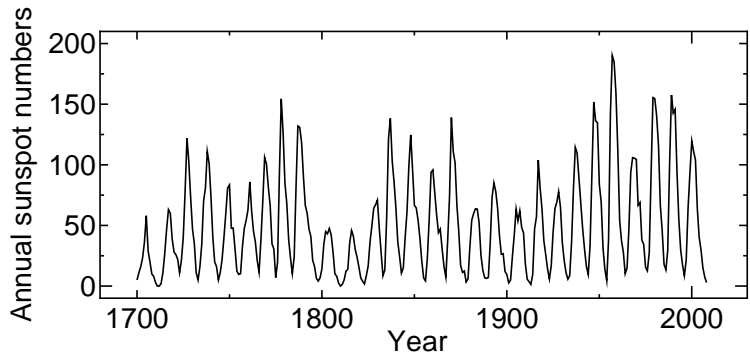}
\end{center}
\caption{The annual sunspot numbers from 1700 to 2008}
    \label{fig:sunspot_numbers}
\end{figure}
\subsection{Annual sunspot numbers}

\begin{figure}[btp]
\begin{center}
\includegraphics[clip, width=8.0cm]{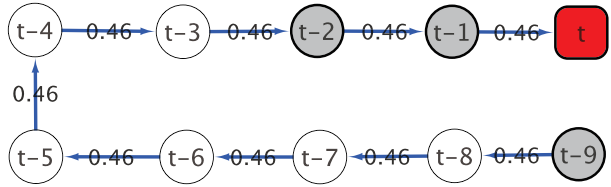}
\end{center}
\caption{(Colour online)~The optimal path network constructed from the model,
         Eq.~(\ref{eq:sunspot}), for the annual sunspot numbers. 
         The nodes are from $ x(t) $ to the largest time delay $ x(t-9) $
         in the model.
         For the explanation of notations used in this figure,
         see Fig.~(\ref{fig:mst network for simple example}).}
 \label{fig:sunspot_network}
\end{figure}

We first apply the proposed method to annual sunspot numbers 
from year 1700 to 2008, and the model is built with time delays 
up to 15~\cite{Judd-Mees:non-uniform_embedding}.
(See Fig.~\ref{fig:sunspot_numbers}.)
The obtained RAR model is 
\begin{align}
     x(t) &= 5.6237 + 1.2108\ x(t-1) - 0.5183\ x(t-2) \notag \\
          &\phantom{=} + 0.2033\ x(t-9) + \varepsilon(t).
 \label{eq:sunspot}
\end{align}
Three time delay terms $ x(t-1) $, $ x(t-2) $ and $ x(t-9) $ are
selected.
and each direct distance to the node~$ x(t) $ is
$ 0.4598 $, $ 2.3689 $ and $ 6.4796 $, respectively.
Figure~\ref{fig:sunspot_network} shows that the optimal path network of 
the nodes from $ x(t) $ to $ x(t-9) $ has a simple chain structure.
This result indicates that the fundamental mechanism of 
the oscillation of annual sunspot numbers is simple and 
the previous sunspot numbers influence the next numbers in series.
From this result we can see the correspondence between 
the periodicities and the chain structure of 
the annual sunspot numbers.

%
\subsection{Microtremor data}

We next apply the proposed method to microtremor data,
which shows rather non-trivial results.
These data were taken from the East-West components of the ground velocity signal
of the 1982 Urakawa-Oki (Hokkaido, Japan) earthquake \cite{Takanami-Kitagawa:book02}.
The data was measured at 50~Hz and consists of 2600 data points in total.

\begin{figure}
\begin{center}
\includegraphics[clip,width=10.0cm]{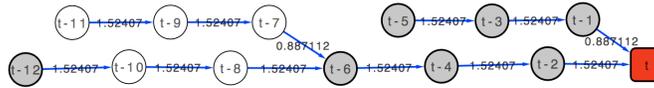}
\end{center}
\caption{
(Colour online)~The microtremor data examined in this paper.
From the entire time interval (0 - 50 sec), three different time regions are
extracted to show the network structure transition indicating the change in the
nature of the microtremor in the course of the oscillation.
The presented optimal path networks are for Region (a) (0 - 12 sec),
Region (b) (20 - 30 sec), and Region (c) (40 - 50 sec).
For other notations used in this figure,
see Fig.~\ref{fig:mst network for simple example}.
Note that the positions of the nodes are irrelevant.
}
	\label{fig:earthquake_data}
\end{figure}

To demonstrate how the network structure changes in the course
of time development of the oscillation, we show in Fig.~\ref{fig:earthquake_data}
the network structures corresponding to three different time regions in this
microtremor data\footnote{The rough boundaries have been 
investigated originally by Kitagawa~\cite{Kitagawa:book10} and we rechecked 
the data and the legitimacy of the boundaries by ourselves.}.
Three models are built with time delays up to 15~\cite{Judd-Mees:non-uniform_embedding}.
It should be noted again that, for the calculation of the direct distances
to the node $ x(t) $, we do not use the constant parameter in the obtained model
and dynamic noise.

%

For the time region~(a), the obtained RAR model has one constant parameter
and eight time delay terms,
\begin{align}
 x(t) &= -0.2203 + 0.3652\ x(t-1) + 0.2811\ x(t-3)   \notag \\
      &\phantom{=} + 0.4054\ x(t-4) + 0.1378\ x(t-6) \notag \\
      &\phantom{=} - 0.1496\ x(t-7) - 0.2024\ x(t-8) \notag \\
      &\phantom{=} + 0.1442\ x(t-11) - 0.1803\ x(t-12) + \varepsilon(t).
    \label{eq:p_wave}
\end{align}
We obtain the following eight direct distances to the node $ x(t) $
corresponding to each term:
$ (1.6859, 2.3422, 1.4552, 5.0961, 4.6778, 3.3920, 4.8630, 3.8415) $.

The network for the time region~(a) in Fig.~\ref{fig:earthquake_data} shows that
while the nodes $ x(t-1) $, $ x(t-3) $, $ x(x-4) $, $ x(t-11) $
and $ x(t-12) $ are directly connected to the node~$ x(t) $,
the nodes $ x(t-6) $, $ x(t-7) $ and $ x(t-8) $ are not.
The network also indicates another interesting feature of this model. 
The region~(a) in Fig.~(\ref{fig:earthquake_data}) implies that
this network structure contains seemingly two ``hubs,''
nodes $ x(t) $ and $ x(t-4) $, because other nodes flock to them.
Two chains of nodes separated by the shortest distance, $1.46$,
with time delay 4 can be found, the one of which
connects nodes $ x(t) $, $ x(t-4) $ and $ x(t-8) $ and the other of which
connects nodes $ x(t-5) $ and $ x(t-9) $.
We can thus consider from this information that time delay~4 is
a major period of this oscillation.
We consider that time delays 1 and 3 also play significant roles that cannot be ignored,
because we can find several chains of nodes with time delay 1 and 3.
For time delay~1, there are the chain of nodes $ x(t) $, $ x(t-1) $ and $ x(t-2) $
and the chain of nodes $ x(t-4) $ and $ x(t-5) $.
For time delay~3, there are the chain of nodes$ x(t-3) $ and $ x(t-6) $
and the chain of nodes$ x(t-7) $ and $ x(t-10) $.
It is highly probable that a hub structure appears when an oscillation
consists of a main oscillation, where each unit oscillation is furthermore affected by
other underlying oscillations with shorter time scales than the main oscillation period.
As the absolute parameter value of the term $ x(t-4) $ is the largest in the model,
one may think that the time difference 4 is the main period
and that the nodes $x(t)$ and $x(t-4)$ can be ``hubs.''
It is not straightforward, however, to know the significance of
time delays 1 and 3 by simply examining the model, Eq.~(\ref{eq:p_wave}).

The RAR model for the time region~(b) is
\begin{align}
 x(t) &= -1.1629 + 1.4154\ x(t-1) - 1.7348\ x(t-2)   \notag \\
      &\phantom{=} + 1.2533\ x(t-3) - 0.8751\ x(t-4) \notag \\
      &\phantom{=} + 0.1473\ x(t-5) - 0.1657\ x(t-7) + \varepsilon(t).
    \label{eq:region_3}
\end{align}
The direct distances to the node $ x(t) $ are
$( 1.6410, 1.2076,\\
   1.9262, 2.9431, 18.4413, 16.3864 )$, respectively.
The RAR model for the time region~(c) is
\begin{align}
 x(t) &= -0.4880 + 0.8847\ x(t-1) - 0.6488\ x(t-2)   \notag \\
      &\phantom{=} + 0.2064\ x(t-3) + 0.2648\ x(t-4) \notag \\
      &\phantom{=} - 0.1846\ x(t-5) + 0.1881\ x(t-6) \notag \\
      &\phantom{=} - 0.1133\ x(t-12) + \varepsilon(t).
    \label{eq:region_5}
\end{align}
The direct distances to the node $ x(t) $ are
$( 0.8871, 1.5241, 5.6417, 4.3525,
6.3287,\\ 6.2076, 10.3854)$, respectively.

From Figs.~\ref{fig:earthquake_data}~(a)-(c),
we notice that the network structure develops
from a star-like structure containing hubs to a relatively simple chain structure
during the course of oscillation.
A chain structure implies that the oscillation at present time is affected
by a previous unit oscillation in a simple manner.
In this way, our network construction method exhibits the transition in
the nature of time series through network structure in addition to
the detailed hierarchical relationship between terms in the model.
One may think that simple visual inspection of
the oscillation profile is sufficient for identifying three regions~(a)-(c).
It should be emphasized, however, that the simple visual inspection
cannot reveal the structural differences in the nature of oscillation
in these regions.

\section{Summary}
\label{sec:summary}

In this paper, we described an algorithm to construct networks
with time structure based on deterministic model structure from time series.
In this algorithm a linear model containing various terms of different time delays 
is transformed into network topology.
The advantage of this method over the existing ones is that 
the global structure of the relationship between terms in a time series model 
is directly translated to the topology of the corresponding network.  
By extracting the optimal paths for the constructed network, 
we can find the global structure and hierarchy between terms.
The transition in the nature of time dependency is represented by
the transformation in the structure of constructed networks.
Using this correspondence between the nature of oscillation
and the structure of the network, we could
compare and categorize different model realizations.
In addition, networks constructed by the proposed method consists of temporal nodes.
It should be emphasized that the insight into the transition in the nature 
of oscillation through network structure described in this paper can only 
be gained by network representation of time series with temporal nodes.
To the best of our knowledge, networks with temporal nodes 
are not well studied and there remains many problems unsolved.
Our arguments and computational examples show the effectiveness of 
introducing networks with time structure and that the proposed method 
has a wide range of applicability and provides us profound insights 
in investigating time dependent phenomena.

The effectiveness and the usefulness of the basic idea presented here
should be verified and enlarged by future works.
At present, we have two following problems.
One is to treat nonlinearity more appropriately and naturally.
This requires to find a proper method for translating nonlinear basis functions
into network topology.
The other is to treat multivariate time series data.
Though we present our method by univariate time series data in this paper,
systems in the real world are not always isolated from their surroundings.
Each factor or element in those systems is interconnected or
interrelated in some way or another to varying degrees.
Hence, it is important to extend our method to be applicable
to multivariate data.

\section{Acknowledgements}

We would like to thank the anonymous referees for their very valuable remarks.
\bibliographystyle{elsarticle-num}

\end{document}